\documentclass[]{aa}

\usepackage{graphicx}

\def\prd{Phys.~Rev.~D} 
\def\apj{ApJ} 
\def\apjs{ApJS} 
\def\mnras{MNRAS} 
\def\aap{A \& A} 
\def\nat{Nat} 

\def\VEV#1{{\left\langle #1 \right\rangle}}

\def\Mll{M_{\ell \ell'}}
\def\Dl{D_\ell}
\def\Cl{C_\ell}
\def\ptchg{260}

\begin{document}

\title{ The CMB temperature power spectrum from an improved analysis of the Archeops data}
\subtitle{}

\author{ 
M. Tristram \inst{1} \and  
G. Patanchon \inst{2} \and
J. F. Mac\'{\i}as--P\'erez  \inst{1} \and  
P. Ade \inst{3} \and  
A. Amblard \inst{4} \and  
R. Ansari \inst{5} \and  
{\'E}. Aubourg \inst{6, \, 7} \and  
A. Beno\^{\i}t \inst{8} \and  
J.--Ph. Bernard \inst{9} \and  
A. Blanchard\inst{10} \and  
J. J. Bock \inst{11, \,12} \and  
F. R. Bouchet \inst{13} \and  
A. Bourrachot \inst{5} \and  
P. Camus \inst{8} \and  
J.-F. Cardoso \inst{14} \and
F. Couchot \inst{5} \and  
P. de Bernardis \inst{15} \and  
J. Delabrouille \inst{7} \and  
F.--X. D\'esert \inst{16} \and  
M. Douspis \inst{10} \and  
L. Dumoulin \inst{17} \and  
Ph. Filliatre \inst{18, \, 7} \and  
P. Fosalba \inst{19} \and
M. Giard \inst{9} \and
Y. Giraud--H\'eraud \inst{7} \and  
R. Gispert \inst{20}$\dag$\thanks{Richard Gispert passed away few weeks
after his return from the early mission to Trapani} \and  
L. Guglielmi \inst{7} \and  
J.--Ch. Hamilton \inst{21} \and  
S. Hanany \inst{22} \and  
S. Henrot--Versill\'e \inst{5} \and  
J. Kaplan \inst{7} \and  
G. Lagache \inst{20} \and  
J.-M. Lamarre \inst{23} \and
A. E. Lange \inst{11} \and  
K. Madet \inst{8} \and  
B. Maffei \inst{3} \and  
Ch. Magneville \inst{6, \, 7} \and
S. Masi \inst{15} \and  
F. Mayet \inst{1} \and  
F. Nati \inst{15} \and  
O. Perdereau \inst{5} \and  
S. Plaszczynski \inst{5} \and 
M. Piat \inst{7} \and  
N. Ponthieu \inst{20} \and  
S. Prunet \inst{13} \and  
C. Renault \inst{1} \and  
C. Rosset \inst{7} \and  
D. Santos \inst{1} \and  
D. Vibert \inst{13} \and  
D. Yvon \inst{6}
}

   \offprints{reprints@archeops.org}

\institute{
Laboratoire de Physique Subatomique et de Cosmologie, 53 Avenue des 
Martyrs, 38026 Grenoble Cedex, France
\and
Department of Physics \& Astronomy, University of British Columbia, Vancouver, Canada
\and
Cardiff University, Physics Department, PO Box 913, 5, The Parade,  
Cardiff, CF24 3YB, UK 
\and
University of California, Berkeley, Dept. of Astronomy, 601
Campbell Hall, Berkeley, CA 94720-3411, U.S.A.
\and
Laboratoire de l'Acc\'el\'erateur Lin\'eaire, BP 34, Campus
Orsay, 91898 Orsay Cedex, France
\and
CEA-CE Saclay, DAPNIA, Service de Physique des Particules,
Bat 141, F-91191 Gif sur Yvette Cedex, France
\and
APC, 11 pl. M. Berthelot, F-75231 Paris Cedex 5, France
\and
Centre de Recherche sur les Tr\`es Basses Temp\'eratures,
BP166, 38042 Grenoble Cedex 9, France
\and
Centre d'\'Etude Spatiale des Rayonnements,
BP 4346, 31028 Toulouse Cedex 4, France
\and
Laboratoire d'Astrophysique de Tarbes Toulouse,
14 Avenue E. Belin, 31400 Toulouse, France
\and
California Institute of Technology, 105-24 Caltech, 1201 East
California Blvd, Pasadena CA 91125, USA
\and
Jet Propulsion Laboratory, 4800 Oak Grove Drive, Pasadena,
California 91109, USA
\and
Institut d'Astrophysique de Paris, 98bis, Boulevard Arago, 75014 Paris,
France
\and
CNRS--ENST
46, rue Barrault, 75634 Paris, France
\and
Gruppo di Cosmologia Sperimentale, Dipart. di Fisica, Univ. 'La
Sapienza', P. A. Moro, 2, 00185 Roma, Italy
\and
Laboratoire d'Astrophysique, Obs. de Grenoble, BP 53,
38041 Grenoble Cedex 9, France
\and
CSNSM--IN2P3, Bât 108, 91405 Orsay Campus, France
\and
CEA-CE Saclay, DAPNIA, Service d'Astrophysique, Bat 709,
F-91191 Gif sur Yvette Cedex, France
\and
Institute for Astronomy, University of Hawaii, 2680 Woodlawn Dr,
Honolulu, HI 96822, USA
\and
Institut d'Astrophysique Spatiale, B\^at. 121, Universit\'e Paris
XI, 91405 Orsay Cedex, France
\and
LPNHE, Universit\'es Paris VI et Paris VII, 4 place
Jussieu, Tour 33, 75252 Paris Cedex 05, France
\and
School of Physics and Astronomy, 116 Church St. S.E., University of
Minnesota, Minneapolis MN 55455, USA
\and
LERMA, Observatoire de Paris, 61 Av. de l'Observatoire, 75014 Paris, France
}

\date{February 16, 2005}

\abstract{ We present improved results on the measurement of the
  angular power spectrum of the Cosmic Microwave Background (CMB)
  temperature anisotropies using the data from the last {\sc Archeops}
  flight. This refined analysis is obtained by using the 6 most
  sensitive photometric pixels in the CMB bands centered at 143 and
  217~GHz and 20~\% of the sky, mostly clear of foregrounds. Using two
  different cross-correlation methods, we obtain very similar results
  for the angular power spectrum. Consistency checks are performed to
  test the robustness of these results paying particular attention to
  the foreground contamination level which remains well below the
  statistical uncertainties. The multipole range from $\ell=10$ to
  $\ell=700$ is covered with 25 bins, confirming strong evidence for a
  plateau at large angular scales (the Sachs--Wolfe plateau) followed
  by two acoustic peaks centered around $\ell=220$ and $\ell=550$
  respectively. These data provide an independent confirmation,
  obtained at different frequencies, of the {\sc WMAP} first year
  results.}
  
\authorrunning{M. Tristram, G. Patanchon, J.F. Mac\'{\i}as--P\'erez et al.}
\titlerunning{Archeops CMB power spectrum}

\maketitle

\section{Introduction}

Observations of the Cosmic Microwave Background (CMB) temperature
anisotropies provide answers to fundamental questions in cosmology.
The experimental determination of the CMB temperature angular power
spectrum (\cite{saskatoon, toco, boom1, maxima1, maxima2, boom2, dasi,
  cbi, vsa, archeops_cl, wmap_cl, barkats, cbi_04, dasi_04}) leads to
important insights into the composition and evolution of the Universe.
Most notable are the conclusions that the geometry of space is
essentially flat, the measurements are consistent with the
inflationary paradigm and the Universe is dominated by unknown forms
of dark energy and dark matter (\cite{line97, macias00, archeops_cosmo,
  douspis03, wmap_cosmo}).

{\sc Archeops}\footnote{see {\tt http://www.archeops.org}} was
designed to obtain a large sky coverage of CMB temperature
anisotropies in a single balloon flight at millimeter and
submillimeter wavelengths. {\sc Archeops} is a precursor to the {\sc
  Planck HFI} instrument (\cite{lamarre}), using the same optical
design and the same technology for the detectors, spider--web
bolometers, and their cooling, 0.1~K dilution fridge. The instrument
consists of a 1.5~m aperture diameter telescope and an array of
21~photometric pixels operating at 4~frequency bands centered at 143,
217, 353 and 545~GHz.  The data were taken during the Arctic night of
February~7,~2002 after the instrument was launched by CNES from the
Esrange base near Kiruna (Sweden). The entire data set covers $\sim
30$\% of the sky.

The {\sc Archeops} initial analysis (\cite{archeops_cl}) -- hereafter
Paper I -- presented for the first time measurements from large
  angular scales to beyond the first acoustic peak ($\ell= 15-350$).
A few months later, the first year {\sc WMAP} results (\cite{wmap})
confirmed the previous measurements and significantly reduced the
error bars on scales down to the second acoustic peak.

This paper presents a second and more refined analysis of the {\sc
  Archeops} data. With respect to Paper I, major improvements on the
timeline processing, the map-making, the beam modeling and the
foreground removal were achieved. Further, new power spectrum
estimation methods based mainly on the cross power spectra between
different detectors maps are used to reduce the contribution from
correlated noise and systematic effects. This essentially allows us to
increase the number of detectors considered (from two for Paper I to
six for this analysis) and to cover a larger fraction of clean sky
(12~\% in Paper I, 20~\% in this paper). These developments lead to a
better sampling and a larger range in multipole space with an improved
accuracy.

The paper is organized as follows: Sect.~\ref{processing} summarizes
the processing on the TOIs (Time Ordered Data) with an emphasis on
changes and improvements with respect to Paper I.
Section~\ref{powerspectrum} describes the methods, Xspect and SMICA,
used for the estimation of the CMB angular power spectrum from
observed emission maps. The estimation of the {\sc Archeops} CMB
angular power spectrum is presented in Sect.~\ref{ccl}. Consistency
checks on the data and the contribution from systematics to the {\sc
  Archeops} CMB angular power spectrum are discussed in
Sect.~\ref{discussions}. A simple comparison with the best-fit
cosmological model provided by the {\sc WMAP} team (\cite{wmap_cosmo})
is shown at the power spectrum level. However, we postpone to a
forthcoming paper the comparison of this dataset to the {\sc WMAP}
data and other datasets at the map level.

\section{Observations and Data processing}\label{processing}
The {\sc Archeops} experiment is described in details in companion
papers. Instrument and data processing are detailed in
\cite{processing}) while the in-flight performances are summarized in
\cite{IFP}. In the following subsections, only key points on the data
processing are summarized and we then focus on refinements implemented
for the present analysis, as compared to Paper I.

\subsection{Observations and standard data processing}\label{dataprocessing}

The instrument contains a bolometric array of 21 photometric pixels,
each one being made of cold optics consisting of an assembly of
back-to-back horns, filters and lenses, and of a 100~mK bolometer,
which operate at frequency bands centered at 143~GHz (8~pixels),
217~GHz (6), 353~GHz (6$=$3 polarized pairs) and 545~GHz (1). The two
low frequencies are dedicated to CMB studies while high frequency
bands are sensitive essentially to interstellar dust and atmospheric
emission. The focal plane is made of 21 spider--web bolometers and
some thermometers and is maintained at a temperature of~$\sim$95~mK by
a $^3$He--$^4$He open--circuit dilution cryostat. Observations are
carried out by spinning the payload around its vertical axis at 2~rpm.
Thus the telescope produces circular scans at a fixed elevation of
$\sim 41$~deg. Observations of a single night cover a large fraction
of the sky as the circular scans drift across the sky due to the
rotation of the Earth and the gondola trajectory.

The {\sc Archeops} experiment was launched on February 7, 2002 by the
CNES\footnote{Centre National d'{\'E}tudes Spatiales, the French
  national space agency} from the balloon base in Esrange, near
Kiruna, Sweden, $68^\circ$N, $20^\circ$E. The night--time scientific
observations span 11~hours of integration.  The pointing
reconstruction, with rms error better than 1~arcmin., is performed
using data from a bore--sight mounted optical star sensor. Each
photometric pixel offset is deduced from Jupiter observations.

Corrupted data (including glitches) in the Time Ordered Information
(TOI), representing less than 1.5\% of the full data set, are flagged.
Low frequency drifts on the data generally correlated to house-keeping
data (altitude, attitude, temperatures, CMB dipole) are removed using
the latter as templates. Furthermore, a high frequency decorrelation
is performed in few chosen time frequency intervals of $\sim$1~Hz
width to remove some bursts of non-stationary high-frequency noise
localised in time and in frequency. The corrected timelines are then
deconvolved from the bolometer time constant and the flagged corrupted
data are replaced by a realization of noise (which is not projected
onto the maps in the map--making step). Finally, low time frequency
atmospheric residuals are subtracted using a destriping procedure
which slightly filters out the sky signal to a maximum of 5\% (see the
red curve on Fig~\ref{figfell}). This effect is corrected for when
computing the CMB angular power spectrum as discussed in Sect.
\ref{debiasing}.

The CMB dipole is the prime calibrator of the instrument. The absolute
calibration error against the dipole as measured by COBE/DMR
(\cite{fixsen}) and confirmed by {\sc WMAP} (\cite{wmap}) is estimated
to be 4\% and 8\% in temperature at 143~GHz and 217~GHz respectively.
These errors are dominated by systematic effects.

As Jupiter is a point-source at the {\sc Archeops} resolution, local
maps of Jupiter allow us to estimate the time constant of the
bolometers and the main beam shape. This is performed using the two
Jupiter observation windows. While the 143~GHz detector beams are
mostly elliptical, the 217~GHz ones are rather irregular (multi--mode
horns). The typical FWHM of the beams is about 12~arcmin. Two Saturn
crossings allowed cross--checks on the time constants and beams.

\subsection{Removal of Galactic and atmospheric foreground emissions}\label{linear_decorr}

The {\sc Archeops} cleaned TOIs at 143 and 217~GHz are contaminated by
atmospheric residuals coming mostly from the inhomogeneous ozone
emission. This contributes mainly at frequencies lower than 2~Hz in
the timeline and follows approximatively a $\nu^2$ law in antenna
temperature. Therefore atmospheric emission is much more important at
the high {\sc Archeops} frequencies (353 and 545~GHz). In the same
way, at the {\sc Archeops} CMB frequencies (143 and 217~GHz) the
Galactic dust emission also contaminates the estimation of the CMB
angular power spectrum even at intermediate Galactic latitudes. Dust
emission, which presents a modified black--body spectrum at about 17~K
with an emissivity of about $\nu^2$, dominates the CMB at high
frequencies and therefore the 353 and 545~GHz channels can be used to
monitor it. To suppress both residual dust and atmospheric signals,
the data are decorrelated using a linear combination of the high
frequency photometric pixels (353 and 545~GHz) and of synthetic dust
timelines.  These are constructed from the extrapolation of IRAS and
COBE observations in the far infrared domain (\cite{SFD,fds}) to the
{\sc Archeops} frequencies. We actually construct a synthetic dust
template for the considered CMB bolometers and also for the high
frequency bolometers so that we can take into account simultaneously
in such a model both types of frequency behaviors.

\begin{figure}
  \resizebox{\hsize}{!}{\includegraphics[scale=0.5]{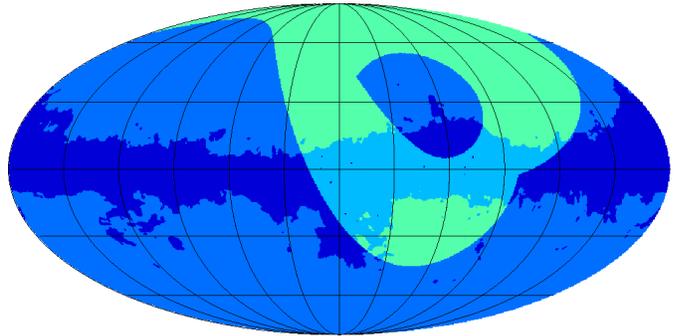}}
  \caption{
    Galactic mask (dark blue lane) applied to the {\sc Archeops}
    coverage (annular green region). The CMB mask is obtained by
    requesting the SFD brightness at 353~GHz to be $<
    0.5$~MJy.sr$^{-1}$. The Mollweide projection of the celestial
    sphere is in Galactic coordinates centered on the Galactic
    anti--center. Gridding on the full sky map is by 30~degree steps.
    The CMB analysis includes 20~\% of the sky (dark green area) while
    {\sc Archeops} covers $\sim$30~\% of the sky. The previous
    analysis only covered 12~\% of the sky above the 30~degree
    Northern parallel.}
\label{figmask}
\end{figure}

As the decorrelation is not perfect in the Galactic plane, a Galactic
mask is then applied to the {\sc Archeops} maps for determination of
the CMB power spectrum. This mask is deduced from a Galactic dust
emission model (\cite{SFD,fds}) at 353~GHz. The Galactic plane and the
Taurus region are efficiently masked by considering only regions with
a brightness \mbox{$< 0.5$~MJy.sr$^{-1}$}. Applying this mask, the CMB
maps derived from the {\sc Archeops} data cover 20~\% of the sky
sampled by $\sim 100,000$ pixels of 7~arcmin. (HEALPix $nside=512$).
Figure~\ref{figmask} presents the {\sc Archeops} coverage to which we
have superimposed the Galactic mask. Only the Northern part above
30~degree was used in Paper I.

\subsection{Map-making}\label{mapmaking}

The noise power spectrum of the {\sc Archeops} TOIs is nearly flat
with increasing power at very low time frequencies due to residuals
from atmospheric noise, and at very high time frequencies due to the
deconvolution from the bolometer time constants. To cope with these
two features on the {\sc Archeops} noise we have used an optimal ({\it
  i.e.} it achieves least square error on pixelised map) procedure
called MIRAGE (\cite{mirage}) to produce maps for each of the
detectors.

MIRAGE is based on a two-phase iterative algorithm, involving optimal
map-making together with low frequency drift removal and Butterworth
high-pass filtering. A conjugate gradient method is used for resolving
the linear system. A very convenient feature of MIRAGE is that it
handles classic experimental issues, such as corrupted samples in the
data stream, bright sources and Galaxy ringing effects in the
filtering and in the calculation of the noise correlation matrix.

Maps are computed with 7~arcmin. pixels (HEALPix $nside=512$) for each
absolutely calibrated detector with their data time band--passed
between 0.1 and 38~Hz. This corresponds to about 90~deg. and
20~arcmin. scales, respectively. The high--pass filter removes
remaining atmospheric and Galactic contamination. The low--pass filter
suppresses non--stationary high frequency noise.

\begin{figure}
  \resizebox{\hsize}{!}{\includegraphics{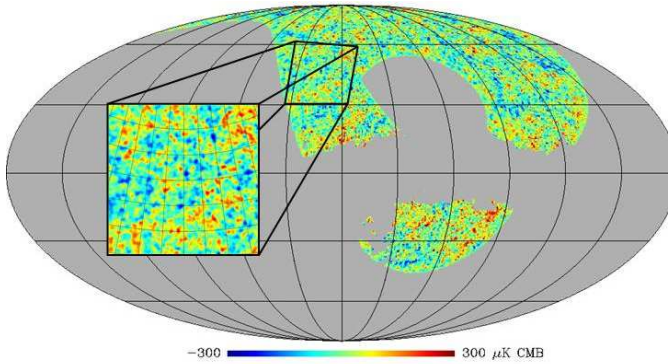}}
  \caption{
    {\sc Archeops} map of the CMB sky in Galactic coordinates centered
    on the Galactic anti-center after smoothing with a 30~arcmin.
    Gaussian. A patch of the sky of 30$\times$30~deg., with high
    redundancy and centered on ($l$, $b$) = (195,45) degrees is zoomed
    up. Gridding on the full sky map is by 30~degree steps, gridding
    on the zoomed patch is 5~deg. The Galaxy is masked as described
    above.}
  \label{mapcmb}
\end{figure}

About two thirds of the {\sc Archeops} sky are observed with $\sim$20
to 60~samples per bolometer and per square degree and one third with a
higher redundancy, about~75 samples per bolometer and per square
degree. For illustration, Fig.~\ref{mapcmb} shows a map obtained from
a weighted linear combination of the maps of each of the six most
sensitive {\sc Archeops} detectors. This map is smoothed with a
30~arcmin. gaussian beam and has a typical rms noise of 50~$\mu$K per
30~arcmin pixel.

\section{Power spectrum estimation}\label{powerspectrum}

In this section, we present three methods, Xspect (\cite{xspect}),
SMICA (\cite{patanchon03}) and power spectrum on the rings
($\Gamma_{m}$ hereafter) (\cite{ansari}) used for the determination of
the angular power spectrum of the CMB temperature anisotropies with
the {\sc Archeops} data. Beforehand we detail the procedure we use to
correct from beam smoothing and filtering effects as well as from
inhomogeneous coverage.

We have thoroughly probed Xspect and SMICA with simulations which are
described below. Results from both methods are included in this paper
to cross validate the final results.  The $\Gamma_{m}$ method is
provided here to illustrate its potential in the estimation of the
angular power spectrum directly from ring data and is more suitable to
Planck-like data.

Xspect and SMICA are based on the so-called `pseudo-$\Cl$'s estimators
(\cite{peebles,spice,master}) which directly compute the pseudo power
spectrum from the spherical harmonics decomposition of the maps. These
spectra are then corrected from the sky coverage, beam smoothing, data
filtering, pixel weighting and noise biases.

A pseudo power spectrum $D_\ell$ is linked to the true power spectrum
$\Cl$ by
\begin{equation}
  \widehat{D_\ell}
  =
  \sum_{\ell'} \Mll p_{\ell'}^2 B_{\ell'}^2 T_{\ell'}
  \VEV{C_{\ell'}}+\VEV{N_\ell} \, \, \, .
  \label{pseudo}
\end{equation}
where $\Mll$ is the mode-mode coupling matrix, $B_\ell$ is the beam
transfer function describing the beam smoothing effect, $p_\ell$ is
the transfer function of the pixelization scheme of the map describing
the effect of smoothing due to the finite pixel size and geometry,
$T_\ell$ is an effective transfer function that represents any
filtering applied to the time ordered data, and $\VEV{N_\ell}$ is the
noise power spectrum.

In the following, the $\Mll$ matrix describes the mode-mode coupling
resulting from the incomplete sky coverage and the weighting applied
to the sky maps. We take into account the $p_\ell$ pixel transfer
function due to the smoothing effect induced by the finite size of the
map pixels. This function is provided in the HEALPix package
(\cite{healpix}).

\subsection{Beam smoothing effect}\label{secbell}

\begin{figure}
  \resizebox{\hsize}{!}{\includegraphics{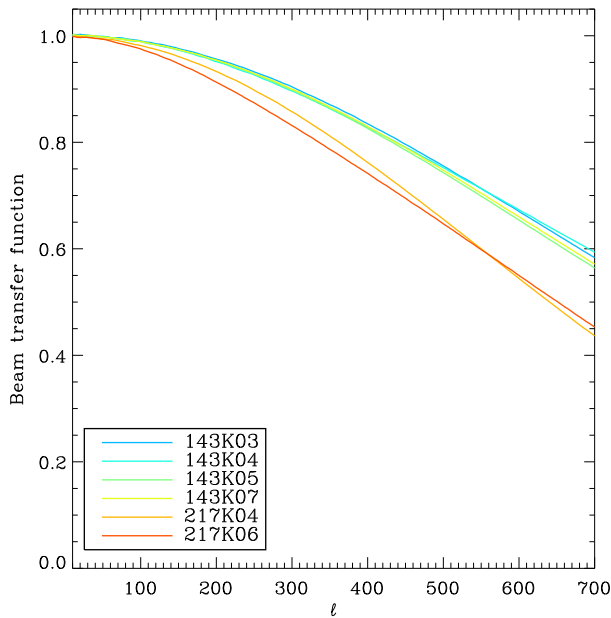}}
  \caption{
    Beam transfer functions of the six most sensitive {\sc Archeops}
    detectors computed using the {\it Asymfast} beam description.}
  \label{figbell}
\end{figure}

Most of the beams of the {\sc Archeops} detectors have been measured
on Jupiter to be elliptical. A few of them are irregular. Therefore,
the effective beam transfer function must be carefully estimated for
each bolometer. The beam transfer functions are computed from
simulations using the {\it Asymfast} method detailed in
(\cite{asymfast}). This method is based on the decomposition of the
beam into a sum of Gaussians for which convolution is easy in the
spherical harmonic space (up to 12 Gaussians are used here). This
allows us to deal with asymmetric beam patterns using the scanning
strategy of the instrument. Figure~\ref{figbell} shows the beam
transfer function for each of the {\sc Archeops} detectors used in
this analysis. They are estimated with a Monte-Carlo of 100~{\it
  Asymfast} simulations per bolometer. The beam transfer functions for
the 143~GHz detectors are very similar and close to circular Gaussian.
The 217~GHz detector beams are larger and more irregular, and
smear-out more the high multipoles.

The Asymfast method produces negligible ($<$ 0.1 \%) statistical
uncertainties on the $B_{\ell}$ estimation. However, as the beam
patterns have been measured on Jupiter maps they may differ from the
effective beams on the CMB anisotropies. This comes mainly from
uncertainties on the electromagnetic spectral dependence, far-side
lobes, baseline subtraction and time constants, each of which
estimated to be lower than 5\%. For such systematics it is difficult
to estimate their impact on the beam transfer function. As an
illustration, we give conservative upper limits on the $B_{\ell}$
uncertainties by taking, as 1-sigma level error, a third of the
difference between resulting transfer function from elliptical beams
(\cite{fosalba}) and that from the Asymfast decomposition in multiple
Gaussians.  Fig.~\ref{figsyste} shows the uncertainties on the
$C_\ell$s due to the beam transfer function uncertainties. They are
well below the statistical error bars.

\subsection{Filtering and inhomogeneous coverage effects}\label{debiasing}

Filtering leads to a preferred direction on the sky (the scanning
direction) and so the assumption of isotropic temperature fluctuations
implicitly done in Eq.~\ref{pseudo} is not valid any more.  However,
to a first approximation, the bias on the CMB power spectrum due to
the filtering of the time ordered data can be accounted for in the
spherical harmonic space through the $T_\ell$ transfer function.

For this analysis we have performed two types of filtering associated
with the destriping of the data discussed in
Sect.~\ref{dataprocessing} and with the band-pass filter applied to
the data on the map making procedure.

The band-pass filter function $F_\ell$ is computed from 100
simulations of the CMB sky. The simulated maps are converted into
timelines using {\sc Archeops} pointing. These timelines are then
filtered as the {\sc Archeops} data. Subsequently, they are projected
onto maps and the power spectrum of those is compared to the power
spectrum obtained from maps of the same but unfiltered timelines.

\begin{figure}
  \resizebox{\hsize}{!}{\includegraphics{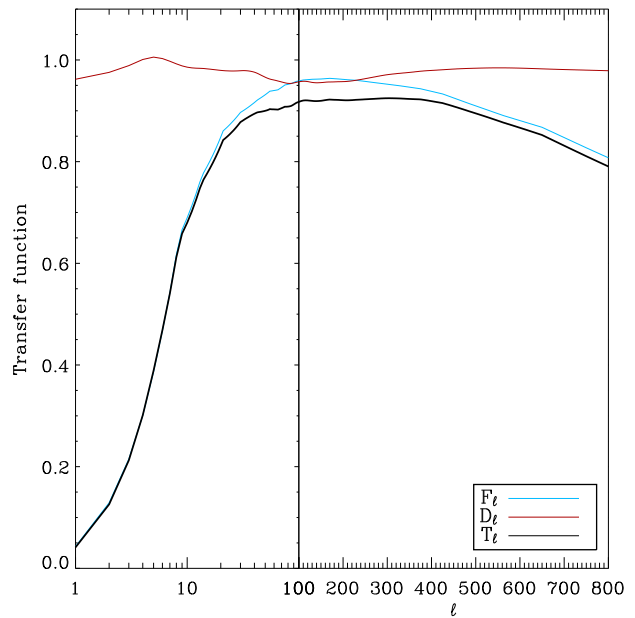}}
  \caption{\label{figfell}
    {\sc Archeops} transfer functions: $F_\ell$ filter function (in
    blue), $D_\ell$ destriping transfer function (in red) and $T_\ell$
    total {\sc Archeops} transfer function (in black).}
\end{figure}

Figure~\ref{figfell} shows in blue the band-pass filter function. It
reaches 65~\% at $\ell = 10$ and remains above 85~\% in the multipole
range [25--700]. In our analysis, all bolometers are identically
filtered and the difference between their pointing vectors is very
small as these bolometers are distributed onto two rows separated by
only $\sim$30~arcmin in the focal plane. We therefore assume an
identical $F_\ell$ function for all detectors. Uncertainties on the
estimation of the $F_\ell$ function are derived from the dispersion of
the simulations.

The transfer function associated with the destriping, $D_\ell$, has
been computed using simulations and is shown in red on
Fig.~\ref{figfell}. The accurate determination of this function is
difficult because the destriping procedure is non linear and CPU
intensive. Thus, in order to be very conservative, we choose to take a
third of the estimate of the function itself as the systematic error
for it.

The total transfer function used for the {\sc Archeops} pipeline
$T_\ell = F_\ell \times D_\ell$ is plotted in black on
Fig.~\ref{figfell}. The uncertainties on the final power spectrum due
to the errors on the $T_\ell$ function are represented on
Fig.~\ref{figsyste}.

\subsection{Xspect}\label{xspect}

The {\sc Archeops} angular power spectrum has been computed using an
extension of the `pseudo-$\Cl$' method to cross power spectra called
Xspect (\cite{xspect}). Assuming no noise cross--correlation between
different detectors, the noise term in Eq.~\ref{pseudo} vanishes and
each cross power spectrum, $A \neq B$, is an unbiased estimate of the
$\Cl$s. Pseudo cross power spectra can be easily corrected from
inhomogeneous sky coverage, beam smoothing and filtering effects by
extending Eq.~\ref{pseudo} into:

\begin{equation}
  \widehat{\Dl^{AB}}
  =
  \sum_{\ell'} \Mll^{AB} p_{\ell'}^2 B_{\ell'}^A B_{\ell'}^B T_\ell
  \VEV{C_{\ell'}^{AB}}
  \label{pseudo_cross}
\end{equation}
where the beam transfer functions $B_\ell$ for each bolometer and the
transfert function $T_\ell $ are those previously described. The
mode-mode coupling kernel $\Mll^{AB}$ is computed for each cross power
spectra from the cross-power spectrum of the weighted masks.  For the
noise weighting scheme we consider a different noise weighted mask for
each {\sc Archeops} detector.  This mask is constructed by multiplying
the mask on Fig.~\ref{figmask} by the inverse of the noise variance on
each pixel and is convolved by a 30 arcmin.  Gaussian.

After correction, all cross power spectra $\VEV{C_{\ell'}^{AB}}$ are
combined into a single estimate of the power spectrum,
$\widetilde{\Cl}$, by weighted averaging assuming the correlation
between multipoles to be negligible. This last assumption is not
completely true, as we can see some correlation at low multipoles on
Fig.~\ref{figcovmat}. Thus the estimate is not completely optimal but
no measurable bias has been found in tests of Xspect on realistic
simulations of Archeops data sets. Analytical estimates of the
covariance matrix and of the error bars in the power spectrum are also
given.

Xspect is designed to estimate both the angular power spectrum and its
error bars even with incomplete sky coverage and mask inhomogeneities,
as is the case with the present {\sc Archeops} data. The approach has
been validated with simulations including realistic noise and CMB
temperature anisotropies. The noise timelines are simulated from an
estimation of the Fourier power spectrum of the noise (\cite{amblard})
for each of the photometric pixels. The CMB signal is simulated using
the HEALPix software from the {\sc Archeops} best-fit $\Lambda$CDM
model (\cite{archeops_cosmo}) convolved by the beam transfer function.
Signal and noise are added into a single timeline which is filtered as
the {\sc Archeops} data and projected on the sky using the {\sc
  Archeops} pointing.

Three sets of 1000 simulations have been computed for sky maps with
HEALPix resolution $nside = 512$~: a first one using an uniform
weighting, a second one using a noise weighting scheme, and a third
one with no noise added. Simulations were performed using the same
optimal map-making method (\cite{mirage}) as the one used for the
data.

From these simulations we have found that there is no bias at the 1\%
level in the estimation of the power spectrum. The analytical error
bars provided by Xspect are also found to be above the standard
deviation in the simulations by less than 10\% and with a rms of 7\%.
Moreover, the noise contribution to the error bars on the simulated
data and the {\sc Archeops} data are in agreement within 5\%.
Hereafter, we will use the analytical estimates provided by Xspect for
the error bars of the {\sc Archeops} angular power spectrum excluding
the sample variance contribution. The latter is computed from the
dispersion of the simulations without noise and is added up to obtain
the final error bars on the CMB angular power spectrum. Therefore, the
sample variance contribution to the error bars is given by the
best-fit {\sc Archeops} model described in \cite{archeops_cosmo}.
 
As mentioned ealier, an improvement of about 10\% on the error bars is
obtained by using uniform weighting at low multipoles and a noise
weighting scheme at high multipoles. Thus, in the following all power
spectra presented are computed using uniform weighting up to $\ell <
\ptchg$ and using a noise weighting scheme for $\ell \ge \ptchg$.

\subsection{SMICA}\label{smicasection}

Using the filtering and beam transfer functions as well as the masks
described in Sect.~\ref{secbell} and \ref{debiasing}, we process the
Archeops maps with a different estimation method of the CMB angular
power spectrum: SMICA (Spectral Matching Independent Component
Analysis) (\cite{patanchon03}).

A specificity of SMICA is its ability to estimate jointly the power
spectra of several underlying components (including noise) assuming
that the observed sky is a linear combination of components. In
spherical harmonic space and in a matrix form, the model is~:
\begin{equation}
  x_{\ell m} = A~s_{\ell m} + n_{\ell m}
  \label{modelSMICA}
\end{equation}
where $x_{\ell m}$ is a vector of spherical harmonics coefficients of
the observed maps for each of the considered detectors; $A$ is the
$N_d$ (number of detectors) $\times$ $N_c$ (number of components)
mixing matrix which defines the amplitude of the different components
in each observed map. The coefficients of $A$ are related to the
electromagnetic spectra of the components and to the relative
calibration between detectors. The spherical harmonic coefficients of
the components and noise are stored in vectors $s_{\ell m}$ and
$n_{\ell m}$.

SMICA is based on matching empirical auto- and cross spectra to their
expected forms, as predicted by model~(\ref{modelSMICA}) and by the
statistical assumption of decorrelation between components. The
mismatch is measured by a measure of divergence between the measured
and modeled spectra which stems from the likelihood of a Gaussian
stationary model. The adjustable parameters are: the power spectrum of
each of the components (including CMB and noise) as well as the mixing
matrix $A$. A complete description of SMICA is given in
\cite{delabrouille,cardoso,patanchon03}.

In the specific case of Archeops, spectral statistics are formed as
follows. The spherical harmonic coefficients $x_{\ell m}$ are computed
on the sky region which is common to all detectors using two different
weighting schemes. For $\ell < \ptchg$, pixels are uniformly weighted.
For $\ell \ge \ptchg$, pixels are weighted proportionally to the
number of data samples per pixel for the best detector. Band-averaged
pseudo auto- and cross-power spectra are formed from these $x_{\ell
  m}$ and corrected for beam smoothing.  If $Q$ bands are used, we
obtain in this manner a set of $Q$ spectral matrices $\widehat R_q$
($q=1,\ldots, Q$), each of size $N_d\times N_d$.  Next, we choose
which parameters should be be estimated (power spectra for CMB and
possibly other components, all or parts of the coefficients, noise
levels), collect all these parameters into a vector $\theta$ and
denote $R_q(\theta) = \VEV{\widehat R_q}$ the expected value of the
spectral matrices for a given value of $\theta$ (this is easily
computed from model~(\ref{modelSMICA})). The SMICA algorithm estimates
the unknown parameters by minimizing the spectral mismatch
\begin{equation}
  \phi(\theta) = \sum_q w_q K( \widehat R_q , R_q(\theta))
  \label{eq:defmisfit}
\end{equation}
where $w_q$ is the number of independent $a_{lm}$ in the $q$th
spectral band and where the mismatch measure $K(\cdot,\cdot)$ between
two positive matrices is defined as $K(M_a, M_b) =
\frac12\left(\mathrm{trace}(M_aM_b^{-1}) - \log\det M_aM_b^{-1}
  -N_d\right)$ (with this choice, the estimated parameter
$\widehat\theta=\arg\min\phi(\theta)$ is a maximum likelihood estimate
as shown in~\cite{delabrouille}). The resulting estimated power
spectra are then corrected from partial coverage and filtering effects
using the MASTER formalism described in Sect.~\ref{debiasing}.

In order to evaluate error bars and possible biases, we have performed
500 realistic simulations of {\sc Archeops} data. The data model
includes synthetic CMB emission (observed with the same scanning
strategy as used by \textsc{Archeops}) and noise for each detector.
Application of SMICA to these simulated data has not shown any
measurable bias.

Error bars for the estimated power spectra can also be obtained
analytically from the Fisher information matrix. They have been
compared to the dispersion found in the Monte-Carlo simulations.
Analytic error bars on the CMB power spectrum are found to be slightly
underestimated (about 10\% on average). In the following, we use the
analytic error bars corrected from the factor measured in the
simulations.

\subsection{CMB power spectrum on the rings}\label{gammaM}

A third approach based on one-dimensional properties of the CMB
inhomogeneities on rings has been performed on {\sc Archeops} data
(\cite{ansari,plas}). It has been made possible by the {\sc Archeops}
sky scanning strategy, which scans quasi circles on the sky. The fact
that we directly use TOI information with no requirement of projection
on maps of the sky makes this method complementary to the two previous
ones.

$\Gamma_m$ is defined as the Fourier power spectrum of the signal on a
sky ring. For a ring of colatitude $\Theta$, the relation between
$\Gamma_m(\Theta)$ and the $\Cl$ (\cite{del}) follows:
\begin{eqnarray}
\label{gmtocl}
\Gamma_m(\Theta) &=& \sum_{\ell=|m|}^\infty \Cl T_\ell 
B_\ell^2 {\cal(P)}_{\ell m}^2(\Theta) \ .
\end{eqnarray}
where $T_\ell$ is the transfer function for the destriping and
filtering, $B_\ell$ is the beam transfer function and ${\cal(P)}_{\ell
  m}$ are the Legendre polynomials.

Rings are built for each bolometer from the TOIs by using the pointing
information. They are then analysed by pairs. For each ring pair
($i-1$,$i$), whenever measurements taken at the same angular phase
$\phi$ are separated on the sky by less than $0.1$~degree, we define a
``signal'' $S_i(\phi)$ and a ``noise'' $N_i(\phi)$ as respectively the
half sum and half difference of the measurements from each rings.

Once these quantities are computed ring per ring, we analyse $S$ and
$N$ in two ways. On the first hand, we compute the difference of the
mean values of their Fourier spectra (that we call the $\Gamma_m$
analysis). On the other hand, the average of the auto-correlation
functions for each pair is computed and then Fourier transformed to
obtain the $\Gamma_m$ power spectrum. In both cases a Galactic mask
similar to that described in Sect.~\ref{linear_decorr} is applied. In
addition since the autocorrelation approach needs all the low
frequency drifts to be properly removed, we apply a cross-scan
destriping~(\cite{a.bourrachot-thesis}). Since the noise directly
pops-up from the data themselves, no simulation is needed in these
approaches.

The error bars on the $\Gamma_m$ power spectrum are computed from the
dispersion on the Fourier transform across rings and then propagated
to obtain the uncertainties on the angular power spectrum.

\section{Main results}\label{ccl}

\begin{figure}[t]
  \resizebox{\hsize}{!}{\includegraphics{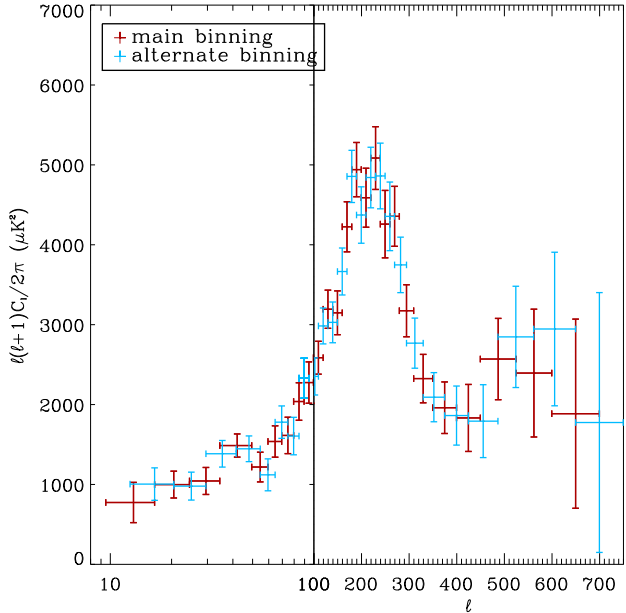}}
  \caption{
    {\sc Archeops} temperature angular power spectrum obtained using
    the Xspect method. A mixing of log-linear scales is presented to
    improve the readibility of the figure both on the Sachs--Wolfe
    plateau and on the acoustic peaks regions. Two intertwined and
    therefore not independent binnings (red and blue) are
    represented.}
  \label{figarchcl}
\end{figure}

The analysis presented in this paper uses the six most sensitive {\sc
  Archeops} bolometers, four at~143~GHz and two at 217~GHz with
instantaneous sensitivities ranging from 93 to 207 $\mu {\mathrm
  K}_{\mathrm CMB} \, {\mathrm s}^{\frac{1}{2}}$. Note that those
instantaneous sensitivities are better, by a factor of at least five,
than those of the {\sc WMAP} satellite mission detectors (\cite{wmap})
and a factor 2 to 4 worse than the nominal ones expected for the {\sc
  Planck--HFI} instrument. We consider 20~\% of the sky by applying
the Galactic mask presented in Sect. \ref{linear_decorr}.

Table~\ref{tab_cl} presents the angular power spectrum measured by
{\sc Archeops}. Results for the Xspect and SMICA methods are both
given as they are based on different assumptions on the data model.

\subsection{{\sc Archeops} temperature angular power spectrum using Xspect}

\begin{figure}
  \resizebox{\hsize}{!}{\includegraphics{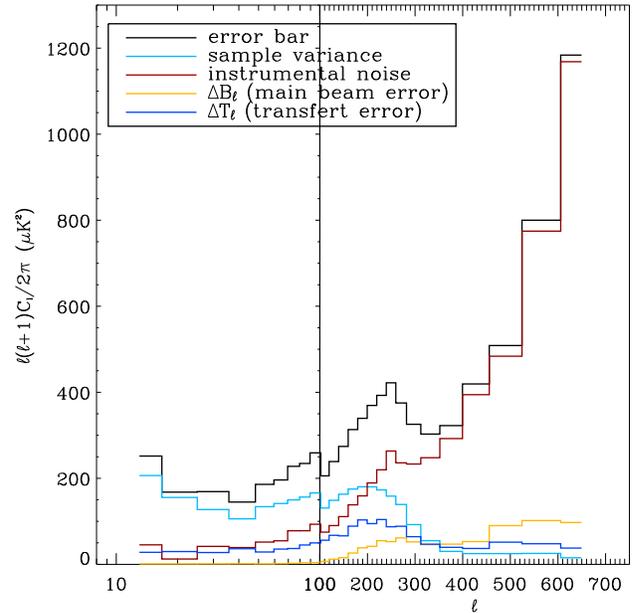}}
  \caption{
    Detailed description of the statistical error bars (in black) on
    the {\sc Archeops} angular power spectrum obtained with Xspect in
    terms of sample variance (in cyan) and instrumental noise (in
    red). In addition, systematic errors on the angular power spectrum
    estimation due to uncertainties on the filter (in blue) and beam
    smoothing function (in yellow) are shown (see
    Sect.~\ref{powerspectrum}).}
  \label{figsyste}
\end{figure}

Figure~\ref{figarchcl} shows the {\sc Archeops} CMB angular power
spectrum obtained using the Xspect method for two intertwined binnings
(blue and red). These binnings correspond to two sets of overlapping
and shifted window functions which lead to two non--independent
estimates of the CMB angular power spectrum. A mix of logarithmic and
linear scales in multipole space is presented to improve the
readibility of the figure both on the Sachs--Wolfe plateau and on the
first two acoustic-peaks clearly detected by {\sc Archeops}.  Two
different weighting schemes are combined to produce the smallest error
bars. At low $\ell$ multipoles a uniform weighting is preferred
whereas for high $\ell$s the sky maps for each detector are noise
weighted by using $w_{p,d}=1/\sigma_{p,d}^2$ where $\sigma_{p,d}^2$ is
the variance of the pixel $p$ of the sky map from the detector $d$.
The two schemes yield identical results around the mixing point, $\ell
\simeq \ptchg$ and they are joined in order to minimize the final
error bars.

\begin{figure}
  \resizebox{\hsize}{!}{\includegraphics[scale=0.5]{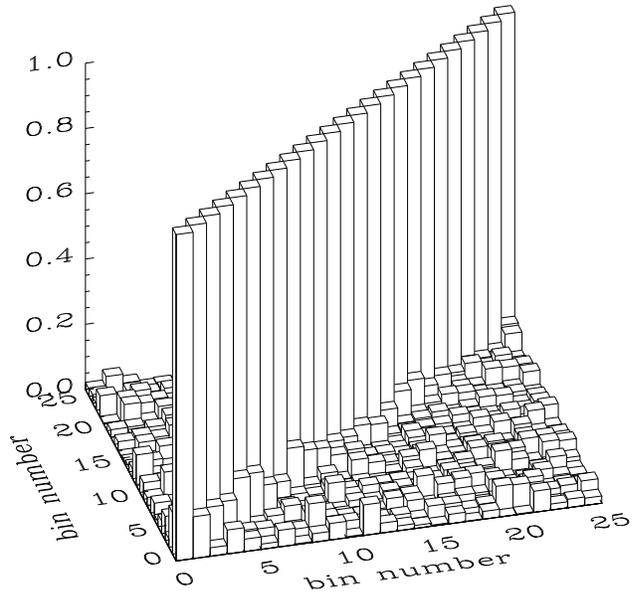}}
  \caption{\label{figcovmat}
    Error covariance matrix of the {\sc Archeops} angular power
    spectrum computed using the Xspect method. The correspondence
    between bin number and multipole range is indicated in
    Tab.~\ref{tab_cl}. The off-diagonal terms are less than 12~\%}
\end{figure}

Figure~\ref{figsyste} shows a detailed description of the statistical
error bars (in black) on the {\sc Archeops} angular power spectrum in
terms of sample variance (in cyan) and instrumental noise (in red).
Sample variance is deduced from the set of simulation without noise.
It corresponds to the uncertainty on the model that is induced by the
fact that we can only look at a part of one realisation of the sky.
Sample variance dominates for $\ell < 100$ and contributes to 50\% or
more of the total statistical error up to $\ell \sim 200$. Systematic
errors due to uncertainties on the filter (in blue) and beam smoothing
function (in yellow), which were computed as discussed in
Sect.~\ref{powerspectrum}, are well below the statistical errors.

Figure~\ref{figcovmat} shows the absolute value of the normalised
error covariance matrix of the {\sc Archeops} angular power spectrum
for the binning shown in red on Fig.~\ref{figarchcl}. The
correspondence between bin number and multipole range is indicated in
Tab.~\ref{tab_cl}. This matrix was computed using the simulations
described in section~\ref{xspect} and provides the absolute
correlation between multipole bins. The off-diagonal terms are less
than 12~\%, and therefore the $\Cl$ estimates can be considered as
roughly uncorrelated across bins on multipole space.

\subsection{{\sc Archeops} temperature angular power spectrum using SMICA}\label{smicapowspec}

To apply the SMICA method to the {\sc Archeops} data we choose to
estimate two components (number required by the data: see
figure~\ref{fitSMICA} and related comments) corresponding to the CMB
anisotropies and to unidentified residuals from foregrounds.  The
mixing matrix is simultaneously estimated allowing for recalibration
of individual detectors against the most sensitive photometric pixel
at 143 GHz.

\begin{figure}
  \resizebox{\hsize}{!}{\includegraphics{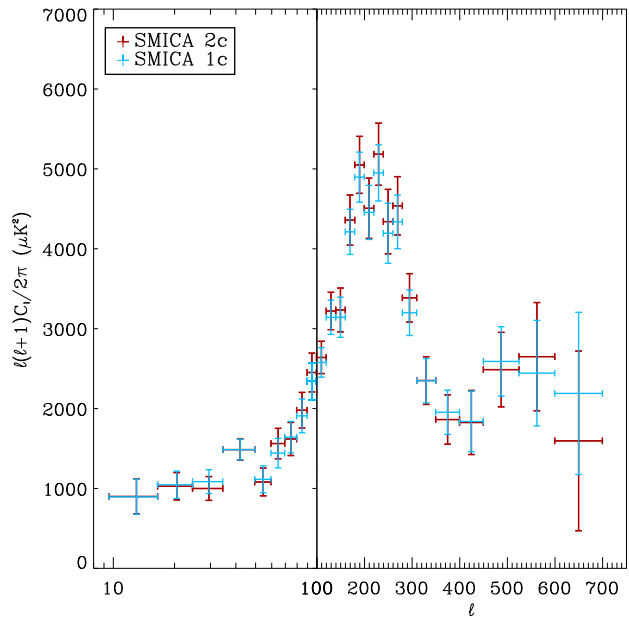}}
  \caption{
    {\sc Archeops} temperature angular power spectrum obtained using
    the SMICA method for one (in red) and two (in blue) components. A
    mixing of log-linear scales is presented to improve the
    readibility of the figure both on the Sachs--Wolfe plateau and on
    the acoustic peaks regions.}
  \label{clsmica}
\end{figure}

We find that CMB anisotropies are clearly detected for all the
bolometers. A second component, much weaker in amplitude, is
significant only in the 217~GHz maps. This component is thought to be
a weak residual of foreground subtraction (see Sect.~\ref{foregrounds}
for a more detailed discussion). Figure~\ref{clsmica} shows in red
the estimated CMB power spectrum with SMICA assuming two components.

To assess the impact of the second component, we run SMICA assuming a
single physical component in the {\sc Archeops} maps, meant to be the
CMB anisotropies. For this second analysis, we fix the CMB mixing
parameters to the values derived from the dipole calibration, allowing
the direct comparison with Xspect. Figure \ref{clsmica} shows in blue
the CMB power spectrum obtained in this way.

The fit of the estimated model to the data is quantified by the lowest
possible value $\phi(\widehat\theta) = \min_\theta\phi(\theta)$ of the
spectral matching criterion Eq.~(\ref{eq:defmisfit}). If the model of
observations is correct (\emph{i.e.} includes the probability
distribution of the data), then $\phi(\widehat\theta)$ should be
statistically small. A finer picture is obtained by splitting the
overall fit of $\phi(\widehat\theta)$ into its components $w_q K(
\widehat R_q , R_q(\widehat\theta))$ as a function of the multipole
bin $q$.  Figure~\ref{fitSMICA} shows the spectral adjustment of the
best one-component model and of the best two-component model.  The
adjustment is much better with two components than with a single
component, indicating that a second component is required by the data.

\begin{figure}
  \resizebox{\hsize}{!}{\includegraphics{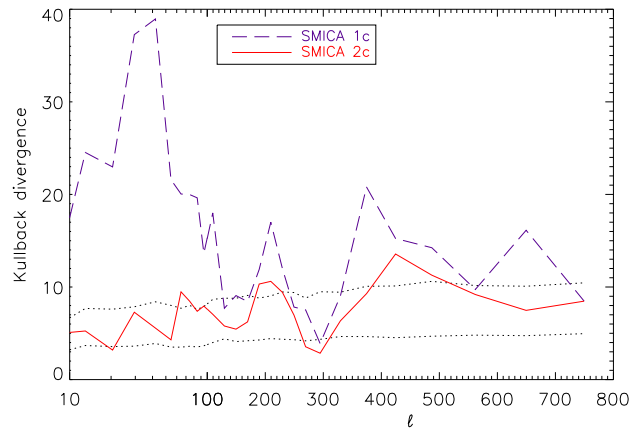}}
  \caption{
    Rescaled spectral fit as function of the multipole. The dashed
    line is the fit for 1~component, the solid line curve is for
    2~components. The dotted lines are the bounds of the 68\%
    confidence interval estimated in simulations of the two-component
    model. Note how a two-component model brings the spectral mismatch
    within the statistical error bounds, showing that in addition to
    the CMB anisotropies a second component is required by the data
    mainly at low spatial frequency ($\ell < 100)$.}
  \label{fitSMICA}
\end{figure}

Blind estimate for two components allows to separate systematic
residuals in the two 217~GHz maps at the cost of some small increase
in the CMB power spectrum error bars. The errors on the estimated CMB
mixing parameters (bolometer intercalibration error) influence the
error bars on the power spectrum estimate. The ratio between CMB power
spectrum statistical error bars for the two and one component cases is
about 20~\% at low $\ell$ and 10~\% at high $\ell$.

\section{Discussion}\label{discussions}

The CMB angular power spectrum measured by Archeops as computed using
Xspect and SMICA extends to a larger multipole range the results
presented in (\cite{archeops_cl}) and is in good agreement with them
on the common multipole range reducing the error bars by a factor of
three.

\subsection{Consistency checks}

\begin{figure}
  \resizebox{\hsize}{!}{\includegraphics{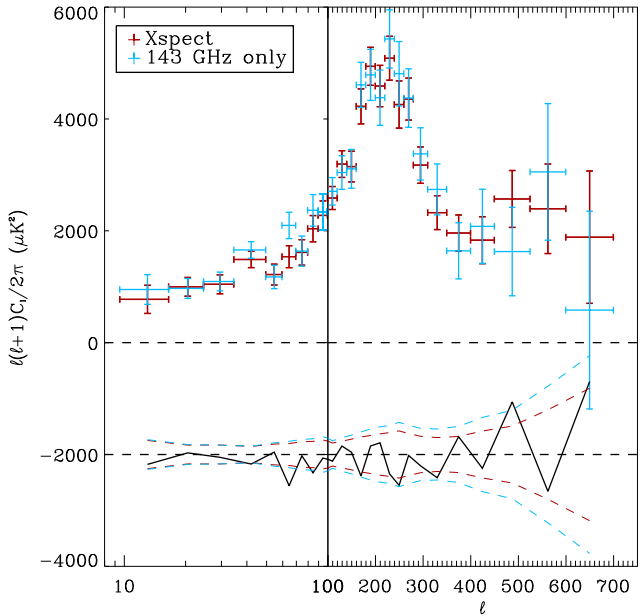}}
  \caption{
    Xspect angular power spectrum using six detectors (in red)
    compared to the one obtained using only the four 143~GHz detectors
    (in blue). The difference between the two power spectra is given
    in the bottom plot (shifted by 2000) and are compared to the error
    bars (black dotted line).}
  \label{cls143}
\end{figure}

Internal tests of consistency have been implemented in order to check
the robustness of the results presented above. The {\sc Archeops} CMB
angular power spectrum has been computed for two different map
resolutions ($nside = 512, 256$ corresponding to 7 and 14 arcmin
pixels resp.) and we observe no significant differences between them.
Furthermore, we have substantially varied the frequency intervals for
the timeline bandpass filtering and no significant effect appears in
the estimation of the angular power spectrum even at high multipoles.
In addition, to check the consistency of the results between the two
CMB channels (143 and 217~GHz) we have computed, using Xspect, the CMB
angular power spectrum for only the four 143~GHz bolometers.
Figure~\ref{cls143} shows this spectrum (in blue) compared to the one
using the 6 most sensitive photometric pixels (in red). The spectra
are in very good agreement, within the error bars, over the full
multipole range. Using only the 143~GHz bolometers reduces
significantly the sensitivity to the second acoustic peak but no
systematic offsets are observed.

\begin{figure}
  \resizebox{\hsize}{!}{\includegraphics{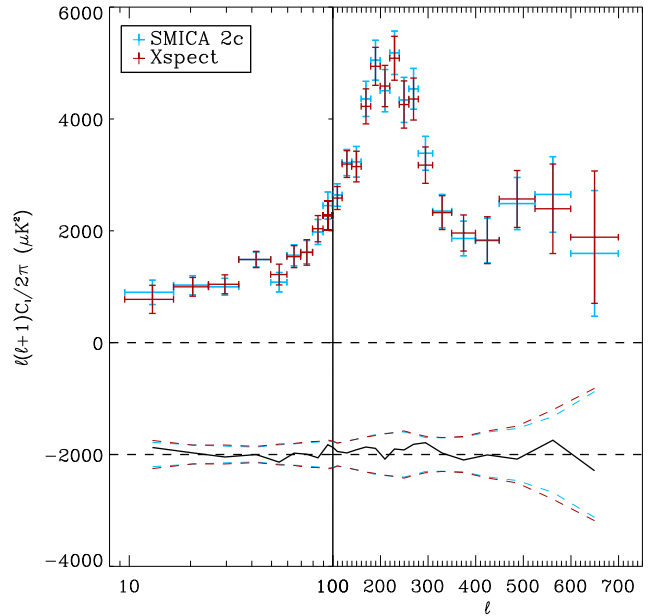}}
  \caption{
    {\sc Archeops} angular power spectrum using Xspect (in red) and
    using SMICA (in blue). The difference between the two power spectra
    is given in the bottom plot (shifted by 2000) and are compared to the error bars
    (blue and red dotted line). See text for details.}
  \label{fig2spect}
\end{figure}

As an extra consistency check, we compare in Fig.~\ref{fig2spect} the
{\sc Archeops} angular power spectrum obtained using Xspect (in red)
with the one computed with 2-components SMICA method (in blue). The
difference between the two power spectra, given in the bottom plot, is
well below the error bars (red and blue dotted line). Detailed
discussion of this issue is presented in Sect.~\ref{foregrounds}.

\subsection{{\sc Archeops} temperature angular power spectrum on the rings}\label{gammapowspec}

\begin{figure}
  \resizebox{\hsize}{!}{\includegraphics{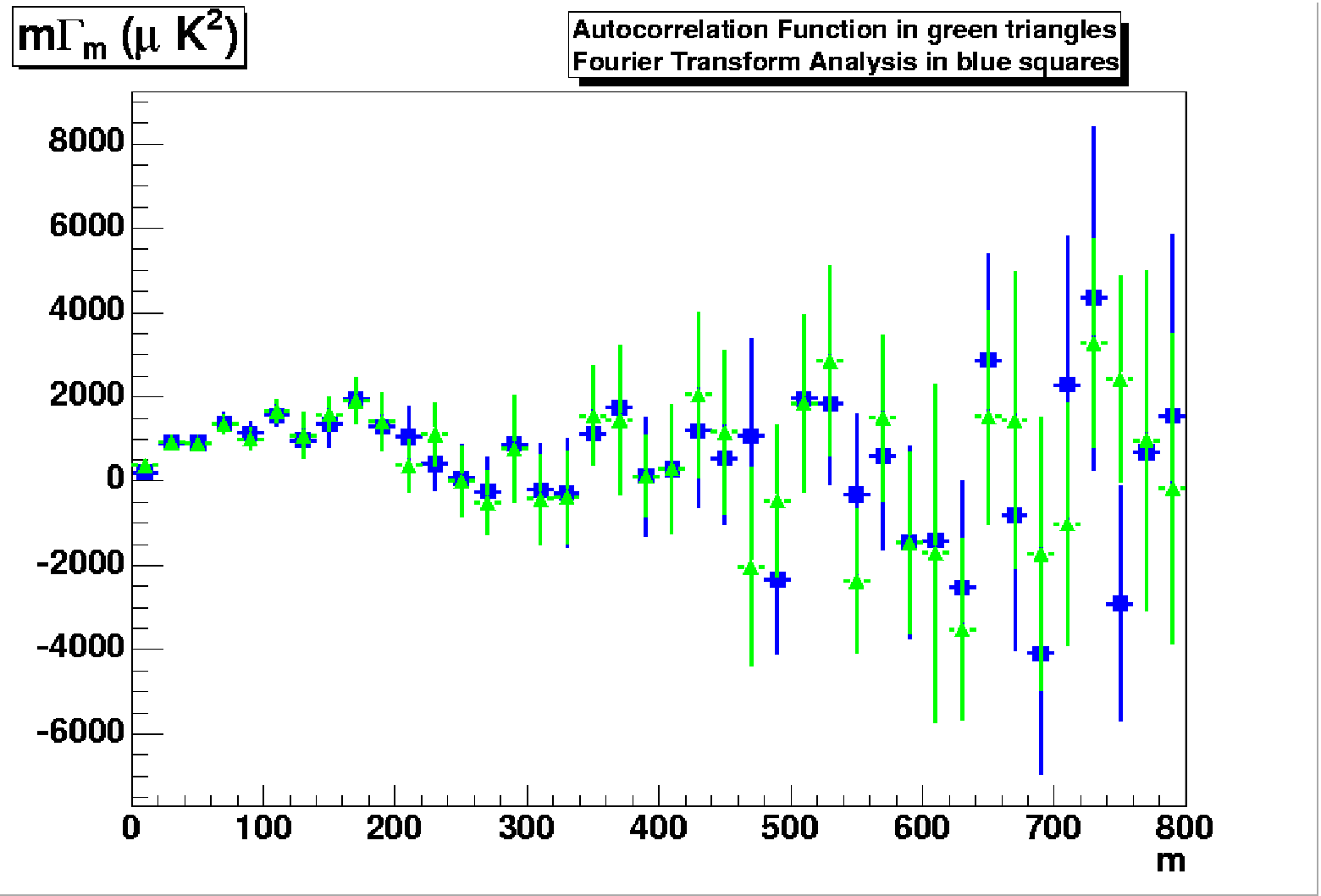}}
  \caption{
    Fourier spectra obtained through the use of the two methods
    described in Sect.~\ref{gammaM} for the best {\sc Archeops}
    bolometer at 143GHz. These analyses are in agreement within the
    error bars.}
  \label{gamma_m}
\end{figure}

We show on Fig.~\ref{gamma_m} the Fourier spectra obtained through the
use of the two ring analysis methods described in Sect.~\ref{gammaM}
for the best {\sc Archeops} bolometer at 143GHz. These analyses are in
agreement within the error bars and show a clear detection of the
first acoustic peak. These results indicate that the processed
timelines contain no obvious spurious feature at a particular time
frequency.

\subsection{Contamination from foregrounds}\label{foregrounds}

As any balloon-borne experiment, {\sc Archeops} is exposed to the
fluctuations of the atmospheric emission. Moreover the Galactic
emission at 143 and particularly at 217~GHz is low but not negligible.
Even if a careful decorrelation to suppress ozone and dust spurious
emissions has been performed (see Sect.~\ref{linear_decorr}), the
residuals from this decorrelation are a potential source of systematic
errors in the determination of the CMB angular power spectrum.

\begin{figure}
  \resizebox{\hsize}{!}{\includegraphics{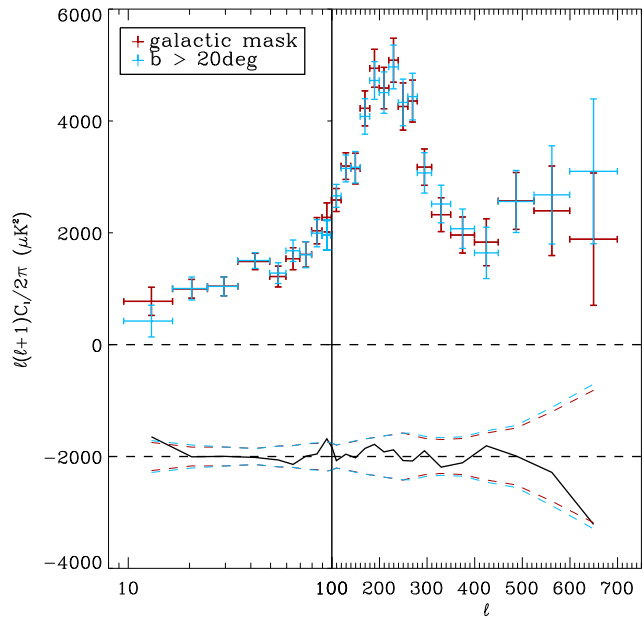}}
  \caption{\label{bcut}
    Xspect {\sc Archeops} power spectrum computed for the Galactic
    mask described in Sect.~\ref{linear_decorr} (in red) and for
    $b>20$ (in blue). The difference between the two estimates is
    given in the bottom plot (shifted by 2000) and compared to the
    error bars (blue and red dotted line).}
\end{figure}

The Galactic dust contribution must be much weaker at high Galactic
latitudes. To assess the level of Galactic residuals, we have computed
the angular power spectrum of the {\sc Archeops} data using only the
Northern part of the {\sc Archeops} sky coverage. Figure~\ref{bcut}
shows the estimate of the angular power spectrum for the Galactic mask
described in Sect.~\ref{linear_decorr} (in red) and for high positive
Galactic latitudes: $b > +20\rm\,deg.$ (in blue) using Xspect. The
differences between the two power spectrum estimates, shown in the
bottom plot, are significantly smaller than the error bars associated
to them. We conclude from this that the residual dust emission in the
CMB angular power spectrum obtained from the {\sc Archeops} data is
small compare to the statistical errors in the multipole range $17 \le
\ell < 700$. The multipole bin $10 \le \ell < 17$ shows a more
important contimation from dust residual emission but still at the
levels of the statistical and systematic uncertainties. For $\ell <
10$ we found that the dust contamination was significant and therefore
this multipole range was not included in this paper. The same test has
been performed using SMICA and leads to identical conclusions.

To fully assess the residual contamination to the {\sc Archeops} data
from dust and atmospheric emissions we have performed two independent
tests based on Xspect and SMICA respectively.

\begin{figure}
  \resizebox{\hsize}{!}{\includegraphics{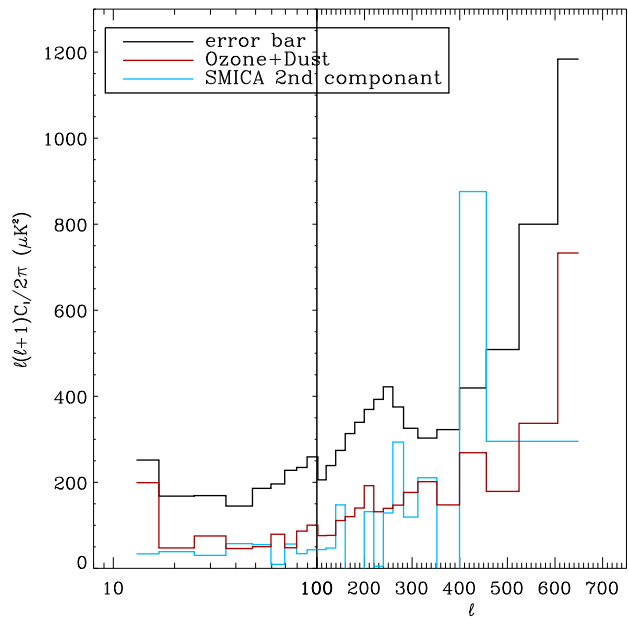}}
  \caption{\label{figforegrounds}
    Atmospheric and Galactic dust residual emissions on the {\sc
      Archeops} 143 and 217~GHz maps. In red, the residual foreground
    emission computed by cross-correlating these maps with the 353~GHz
    maps using the Xspect method. In blue, the residual foreground
    emission obtained from the second component detected by the SMICA
    2~components analysis of the {\sc Archeops} data. In black, we
    plot for comparison the error bars of the {\sc Archeops} CMB
    angular power spectrum.}
\end{figure}

First, using the Xspect method we can cross-correlate the sky maps at
143 and 217~GHz used for the $\Cl$ estimation with the sky maps of the
353~GHz {\sc Archeops} detectors. The observed emission on the latter
is dominated by dust and atmospheric emission and to first order we
can neglect the CMB emission. Thus from this cross correlation, we can
obtain an estimate of the residual foreground contribution to the {\sc
  Archeops} CMB angular power spectrum computed with Xspect. The
results from this analysis are shown on Fig.~\ref{figforegrounds}. The
estimated contamination (in red) remains significantly below the
statistical errors (in black) over the full multipole range except for
the first multipole bin ($\ell$=[10-17[) for which the contamination
is still smaller than the statistical error bar.

As discussed in Sect.~\ref{smicapowspec} we have performed, using
SMICA, a two component analysis of the {\sc Archeops} six best
photometric pixels. The first component on this analysis was
identified as CMB emission whereas the second as the spurious residual
foreground emission. This is significant only for the 217~GHz
bolometer maps. This component is mainly due to residual atmospheric
emission left behind after the linear decorrelation. This estimation
is represented in Fig.~\ref{figforegrounds}, in blue, and can be
compared to the foreground residual contamination estimated with
Xspect at high multipoles. The SMICA estimate is of the same of order
of magnitude and oscillates for $\ell > 200$. These oscillations come
from the uncertainties on the estimation of the second component which
are well reflected on the error bars obtained for it. This could be
due to correlated noise between the 217~GHz maps which would not be
present in the residual foreground estimate obtained using Xspect.
Further, this conclusion is reinforced by the fact that this
contribution does not seem to be fully additive as expected from the
SMICA model.

\begin{figure}
  \resizebox{\hsize}{!}{\includegraphics{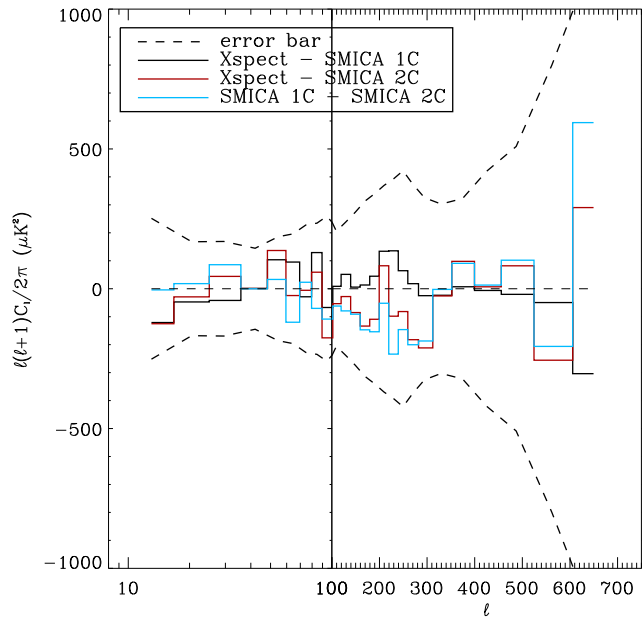}}
  \caption{
    Differences between the {\sc Archeops} CMB angular power spectra
    computed with SMICA 1 and 2 component (in blue), SMICA 1 component
    and Xspect (in black), and SMICA 2 components and Xspect (in red).
    For comparison the statistical error bars are shown (black dashed
    line).}
  \label{diffxspectsmica}
\end{figure}

From the above results we can conclude that the {\sc Archeops} CMB
angular power spectra obtained using Xspect and SMICA are fully
compatible if we take into account the residual atmospheric
contamination which is in any case well below the statistical error
bars as shown in Fig.~\ref{diffxspectsmica}. We have plotted the
differences between the {\sc Archeops} CMB angular power spectra
computed with SMICA 1 and 2 component (in blue), SMICA 1 component and
Xspect (in black), and SMICA 2 components and Xspect (in red). For
comparison the statistical error bars are shown (black dashed line).
This figure visually confirms the fact that the contamination from
foregrounds on the {\sc Archeops} CMB angular power spectrum is well
below the error bars. This analysis of the foreground contamination
validates our choice of the galactic mask described in
Sect.~\ref{linear_decorr}.

Finally, the contribution from point sources is negligible in the
multipole range considered here (see paper I).

\subsection{Comparison to the standard $\Lambda$-CDM cosmological model}

To check the validity of our results and their agreement with previous
cosmological observations we have compared the CMB angular power
spectrum measured by {\sc Archeops} to the best-fit $\Lambda$-CDM
cosmological model presented in (\cite{wmap_cosmo}). This model was
derived from a combination of the {\sc WMAP} data with other finer
scale CMB experiments, ACBAR and CBI and is defined by $h=0.71992$,
$\Omega_{b}h^{2} = 0.02238$, $\Omega_{m}h^{2} = 0.11061$,
$\tau=0.11026$, constant n$_{s}$~(0.05 Mpc)$^{-1}=0.95820$ and
normalization amplitude $A$~(0.05 Mpc)~$= 0.73935$.

\begin{figure}
  \resizebox{\hsize}{!}{\includegraphics{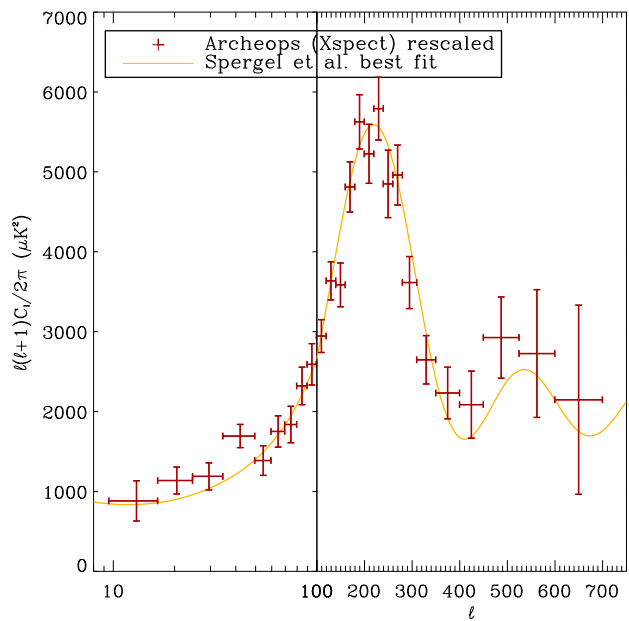}}
  \caption{
    The {\sc Archeops} temperature angular power spectrum rescaled by
    a factor $1.07$ in temperature superimposed on the $\Lambda$-CDM
    best-fit model by the {\sc WMAP} team and presented in
    (\cite{wmap_cosmo}).  }
  \label{figwmap}
\end{figure}

In Fig.~\ref{figwmap} we present the best-fit $\Lambda$-CDM
cosmological model described above superimposed on the {\sc Archeops}
CMB angular power spectrum which is rescaled by a factor $1.07$ in
temperature ($1.14$ in $C_{\ell}$). This factor has been computed by
assuming that the differences between the {\sc Archeops} data and the
model are due to a global scaling factor for all angular scales which
has been fitted to $1.07 \pm 0.02$ with $\chi^2$ of 27/24 and
probability $Q=0.72$. For this fit we have only considered the
statistical error bars on the angular power spectrum.

We observe that the agreement between the rescaled {\sc Archeops} data
and the model is very good. Here the model can be thought of as a
guideline summarising other CMB experiments at different frequencies,
in order to show the overall consistency across the electromagnetic
spectrum.  The scaling factor can be explained by the uncertainties on
the absolute calibration of the {\sc Archeops} data which are 6\% in
temperature (12\% in $C_{\ell}$). A more detailed analysis of this
issue is reported to a forthcoming paper including the determination
of cosmological parameters from the {\sc Archeops} data as well as a
comparison to other CMB observations at the map level.

\section{Conclusion}\label{conclusions}
 
{\sc Archeops} was designed as a test-bench for {\sc
  Planck-HFI\footnote{www.planck-hfi.org}} in terms of detectors,
electronics, cryogenics and data processing. {\sc Archeops} has
demonstrated the validity of these technical choices two years ago by
determining, for the first time and in a single balloon flight, the
temperature angular power spectrum of the CMB from the Sachs--Wolfe
plateau to the first acoustic peak ($15 \le \ell \le 350$) using only
two detectors.

In this paper we present an improved analysis of the {\sc Archeops}
data using the six most sensitive detectors and 20~\% of the sky,
mostly clear of foregrounds. {\sc Archeops} has measured the CMB
angular power spectrum in the multipole range from $\ell=10$ to
$\ell=700$ with 25 bins, confirming strong evidence for a plateau at
large angular scales followed by two acoustic peaks centered around
$\ell=220$ and $\ell=550$ respectively.

The {\sc Archeops} CMB angular power spectrum has been determined
using two different statistical methods, Xspect and SMICA. The results
from these two methods are in very good agreement with differences
between them well below the statistical error bars. Furthermore, they
allow a detailed analysis of the residual foreground contribution
which is mainly due to atmospheric and Galactic dust emissions. The
residual foreground emission on the {\sc Archeops} data is small with
respect to the error bars at all multipoles.

Finally, we have compared the {\sc Archeops} CMB angular power
spectrum to the best-fit $\Lambda$-CDM cosmological model presented in
(\cite{wmap_cosmo}) derived from a combination of the {\sc WMAP} data
with other smaller scale CMB experiments (ACBAR and CBI).  We find
that the {\sc Archeops} data are in very good agreement with this
model considering a rescaling factor to account for uncertainties on
the absolute calibration.

A more detailed analysis for the determination of cosmological
parameter with {\sc Archeops} and other cosmological datasets will be
discussed in a forthcoming paper. Furthermore, a comparison of the
maps from {\sc Archeops}, {\sc WMAP} and other CMB datasets will be
used to study the primordial nature of the measured CMB anisotropies
from their electromagnetic spectrum.

All methods developed for this analysis will be implemented for the
{\sc Planck--HFI} data analysis. Even if {\sc Planck} is less prone to
systematic effects due to its space environment, the know--how
acquired on {\sc Archeops} data should prove useful in order to assess
{\sc Planck} final power spectrum.

\begin{table*}
  \begin{center}
    \begin{tabular}{|ccc|ccc|ccc|c|}\hline\hline
      & & & & & & & & & \\
      & & & & XSPECT & & & SMICA& & \\
      bin & $\ell_{\mathrm{min}}$ & $\ell_{\mathrm{max}}$ &
      $\frac{\ell(\ell+1)}{2\pi}\Cl$  & {\scriptsize total error} &
      {\scriptsize instrumental error} & 
      $\frac{\ell(\ell+1)}{2\pi}\Cl$ & {\scriptsize total error} &
      {\scriptsize instrumental error} & {\scriptsize sample variance} \\
      & & & & & & & & & \\
      \hline
      1 & 10 & 16 &  774 &  251 &   45 &   899 &  217 &   11 & 206\\
      2 & 17 & 24 &  998 &  167 &   12 &  1027 &  170 &   15 & 155\\
      3 & 25 & 34 & 1043 &  168 &   41 &   999 &  149 &   22 & 127\\
      4 & 35 & 49 & 1487 &  144 &   39 &  1486 &  131 &   26 & 105\\
      5 & 50 & 59 & 1217 &  185 &   51 &  1081 &  173 &   39 & 134\\
      6 & 60 & 69 & 1537 &  195 &   54 &  1561 &  189 &   48 & 141\\
      7 & 70 & 79 & 1613 &  227 &   78 &  1619 &  206 &   57 & 149\\
      8 & 80 & 89 & 2038 &  234 &   78 &  1978 &  223 &   67 & 156\\
      9 & 90 & 99 & 2275 &  258 &   93 &  2451 &  242 &   77 & 165\\
      10 &100 &119 & 2586 &  204 &   74 &  2639 &  201 &   71 & 130\\
      11 &120 &139 & 3193 &  238 &   90 &  3221 &  232 &   84 & 148\\
      12 &140 &159 & 3148 &  273 &  110 &  3234 &  274 &  111 & 163\\
      13 &160 &179 & 4225 &  312 &  138 &  4358 &  312 &  138 & 174\\
      14 &180 &199 & 4941 &  339 &  159 &  5050 &  356 &  176 & 180\\
      15 &200 &219 & 4589 &  369 &  189 &  4506 &  377 &  197 & 180\\
      16 &220 &239 & 5085 &  392 &  219 &  5183 &  388 &  215 & 173\\
      17 &240 &259 & 4258 &  421 &  263 &  4340 &  402 &  244 & 158\\
      18 &260 &279 & 4356 &  374 &  235 &  4538 &  365 &  226 & 139\\
      19 &280 &309 & 3174 &  325 &  233 &  3385 &  302 &  210 &  92\\
      20 &310 &349 & 2325 &  302 &  247 &  2351 &  298 &  243 &  55\\
      21 &350 &399 & 1960 &  322 &  292 &  1862 &  309 &  279 &  30\\
      22 &400 &449 & 1832 &  418 &  394 &  1825 &  399 &  375 &  24\\
      23 &450 &524 & 2569 &  507 &  483 &  2487 &  465 &  441 &  24\\
      24 &525 &599 & 2394 &  799 &  774 &  2649 &  676 &  651 &  25\\
      25 &600 &699 & 1885 & 1183 & 1168 &  1595 & 1124 & 1109 &  15\\
      \hline
    \end{tabular}
  \end{center}
  \caption{
    {\sc Archeops} CMB power spectrum and statistical error
    bars (total, instrumental and sample variance) in
    $(\mu\mathrm{K_{CMB}})^2$ computed with Xspect and SMICA (with two
    components) for the best six photometric pixels.}
  \label{tab_cl}
\end{table*}


\begin{acknowledgements}
  We would like to pay tribute to the memory of Pierre Faucon who led
  the CNES team on this successful flight. The HEALPix package was
  used throughout the data analysis~(\cite{healpix}).
\end{acknowledgements}



\end{document}